\documentclass[letterpaper]{article} 
\usepackage{aaai23}  
\usepackage{times}  
\usepackage{helvet}  
\usepackage{courier}  
\usepackage[hyphens]{url}  
\usepackage{graphicx} 
\usepackage{natbib}  
\usepackage{caption} 
\usepackage{algorithm}
\usepackage{algorithmic}

\usepackage{newfloat}
\usepackage{listings}
\DeclareCaptionStyle{ruled}{labelfont=normalfont,labelsep=colon,strut=off} 
\floatstyle{ruled}
\newfloat{listing}{tb}{lst}{}
\floatname{listing}{Listing}
\title{The Half-Life of a Tweet}
\author{
    J\"{u}rgen Pfeffer,
    Daniel Matter,
    Anahit Sargsyan
}
\affiliations{
    School of Social Sciences and Technology\\
    Technical University of Munich\\
    Richard-Wagner-Str. 1, 80333 Munich, Germany
    
}

\usepackage{bibentry}

\begin{document}
\addtolength{\abovedisplayskip}{-2ex}

\maketitle

\begin{abstract}
Twitter has started to share an \emph{impression\_count} variable as part of the available public metrics for every Tweet collected with Twitter's APIs. With the information about how often a particular Tweet has been shown to Twitter users at the time of data collection, we can learn important insights about the dissemination process of a Tweet by measuring its impression count repeatedly over time. With our preliminary\footnote{Data collection and analysis of this short paper are limited, due to the fact that this new feature was released just 10 days before the ICWSM 2023 submission deadline.} analysis, we can show that on average the peak of impressions per second is 72 seconds after a Tweet was sent and that after 24 hours, no relevant number of impressions can be observed for ${\sim}95\%$ of all Tweets. Finally, we estimate that the median half-life of a Tweet, i.e. the time it takes before half of all impressions are created, is about 80 minutes.
\end{abstract}

\section{Introduction}
The idea that information can lose its value over time has long been studied in library science and bibliometrics \cite{Gosnell1944, Burton1960}. A very important metric to assess this value loss is information \emph{half-life}, which describes the time span in which half of the information value is lost. The information value of books can be measured with the number of times a particular book is borrowed from a library, and one way to characterize the value of a scientific article is the number of times an article is cited. Modeling these temporal observations allows us to model decay functions and estimate the time point of 50\% under the curve.

In the context of scientific literature, half-life periods are typically on the time level of years. When we turn to news articles, the half-life in terms of stories published in relation to a certain topic or event comes down to several days. With the advent of the 24-hour news cycle and the rise of social media, the information value of news has suffered an even faster decay \cite{barkemeyer2020media}.

On Twitter, presenting the number of likes and re-Tweets for every Tweet has been an integral part of the platform since its beginning and has been used in order to discuss a wide variety of scores for popularity and to approximate the reach and life span of a Tweet \cite{kobayashi2016tideh, bae2014predicting}. So far, the actual number of how many people have seen a Tweet was only available for a user's own Tweets. 

Starting December 15, 2022, Twitter has been making the number of \emph{views}---which is the name of the \emph{impression\_count} in the platform's GUIs---visible for every Tweet via its web interface as well as via mobile APPs: ``View counts show the total number of times a Tweet has been viewed. With view counts, you can easily see the reach ...''\footnote{https://help.twitter.com/en/using-twitter/view-counts}
On January 5, 2023, it was publicly announced\footnote{\url{https://twitter.com/suhemparack/status/1611085481395224576}}
that the impression count will now also be available via Twitter's API v2 for every Tweet as part of the public metric information. 

\paragraph{Questions and contributions.}
The availability of this feature in the API data has motivated our study. We utilize the Academic API \cite{Pfeffer2023Academic}, which is free for research purposes and allows for full-archive searches on Twitter, We try to answer the following questions, which also enumerate the contributions of this article:

\emph{How can we observe the diffusion dynamics of a Tweet in terms of reach over time?} We will illustrate how the Academic API can be used to repeatedly collect information about the same Tweets in order to create a time series dataset of impressions. 

\emph{What are the properties of the short-term temporal impression distribution, i.e., how many impressions happen when and when is the peak during the early phase?} We show how to use the time series dataset to interpolate an average diffusion curve of impressions on a second timescale. 

\emph{Can we show evidence that the diffusion process of Tweets comes to a relatively early stop so that we have sufficient overall impression counts in order to identify the half-life time points?} We can show that the impression expansion slows down dramatically or even comes to a complete stop for the vast majority of Tweets very quickly so that we can focus our analyses on the first 24 hours of a Tweet's life.

\emph{Finally, can we determine the average half-life of a Tweet?} We will show that, by ignoring a small number of very successful and long-lasting Tweets, we are able to define the median half-life of a Tweet with 79.5 minutes. 

\section{Related Work}
\paragraph{Information half-life of scientific literature.}
Information half-life, i.e., the time it takes until an entity of information has lost half of its value, has been studied for decades in the context of scientific articles \cite{Burton1960} and books in libraries \cite{Gosnell1944}. Historically, half-lives ranging from 3 to 12 years have been observed, with longer half-lives in theoretical sciences \cite{Line1970}. The processes of discovery of new knowledge and the accumulation of existing knowledge underlying the citation process result in the observed half-life phenomenon. Publication delays \cite{Egghe2000Delays} and forgetting knowledge \cite{vanRaan2000} account for some differences in half-life across disciplines. When the half-life of academic material is modeled mathematically, exponential decay functions are used to describe the dynamics \cite{Gosnell1944, Gupta1990, Tsay1998}.

\paragraph{Half-life of news media.}
For news stories, the journalistic production cycles have information decay built into the system as a way to keep readership, viewership \cite{Cushion2021}, and revenues \cite{Clemons2003}. For a specific event that is covered in the news, the half-life is measured as the time until half of the corresponding articles appear. While there are nuanced differences in half-life patterns of media coverage caused by various forms of online and offline external factors \cite{jennings2019street}, the analysis of the dynamics of coverage in printed news outlets reveals a faster decay in light of the emergence of social media \cite{barkemeyer2020media}.

\begin{figure}[b]
\centering
\includegraphics[width=0.95\linewidth]{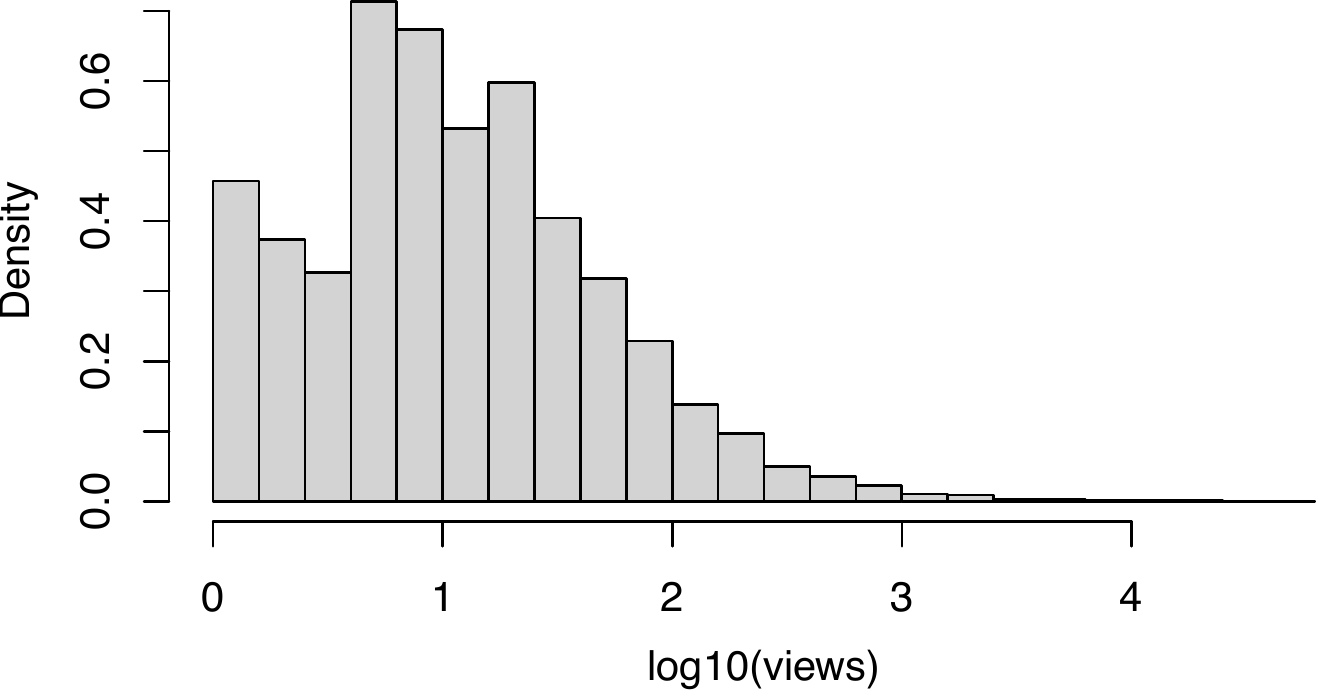}
\caption{Distribution of number of views after ${\sim}30$ Minutes}
\label{fig:views}
\end{figure}

\paragraph{Information decay in social media}
The success of posts on many social media platforms is dependent on shares or views. In the case of Twitter, there are two main approaches to quantify the popularity of a tweet: utilizing the number of retweets, and the audience size, i.e., the number of users who had the tweet in their feed. In the past, one way to calculate the audience size was to use the number of followers for each person who retweets a post. Despite the advantages of potential audience size and of approximation techniques for audience size \cite{kupavskii2013predicting}, the number of retweets, likes and comments have been used in numerous studies to quantify and predict the reach \cite{kobayashi2016tideh} and the lifespan of a tweet \cite{bae2014predicting}. The studies range from analyzing the effect of multimedia on tweet popularity \cite{joseph2018machine, zhao2020comparisons}, the success of personification of brands on Twitter \cite{greene2022brands} to using social media engagement to not only improve predictions of the traffic flow of the news articles, but also to estimate the shelf-life, a variation of half-life, of the articles \cite{ACM_PfefferAlJazeera2014}. 

With the new \emph{impression\_count} variable in Twitter's API data, we are---for the first time---able to directly get information about the reach of a certain Tweet. 

\section{Data}
With Twitter's Academic API v2 \cite{Pfeffer2023Academic}, we have collected 22,144 Tweets on January 6, 2023, as well as the number of views of these tweets in the following way. During the time 9:00-20:00 UTC, we randomly selected ten individual minutes and collected all Tweets (excluding re-Tweets) from the $42^{nd}$ second of these minutes, as described in Pfeffer et al. \citeyearpar{Pfeffer2023Day}. For every second of data (on average 2,214 Tweets), we first started to collect the Tweets exactly 10 seconds after the expiration of the second of interest. After this collection process had finished, we immediately re-started it and collected the same set of Tweets again. We have repeated this collection effort 99 times for every observed second of Twitter data. Since every single API call is limited to a maximum of 500 Tweets, several calls (happening at different timestamps) are necessary for data collection. Consequently, we have stored the exact time of data collection for every API call. For the following analyses, we kept 21,685 Tweets that were available (i.e., not deleted or hidden) in all 99 collection attempts. 

The time series of the Tweet views were, on average, collected over 1,893 seconds (${\sim}31.5$ minutes). While this dataset is sufficient for the majority of the statistical analysis of this article, we do not expect Tweet half-lives to be under half an hour. Consequently, we also collected a second dataset to get a longer period of view data. We performed the following second data collection similar to the approach described above. However, this time we have collected impression counts of about 5,000 Tweets over the course of eleven hours for 1,000 times, as well as the view counts of these Tweets after 24 hours. 


\section{Analyses}

\begin{figure*}[]
\centering
\begin{minipage}{.5\textwidth}
  \includegraphics[width=.9\linewidth]{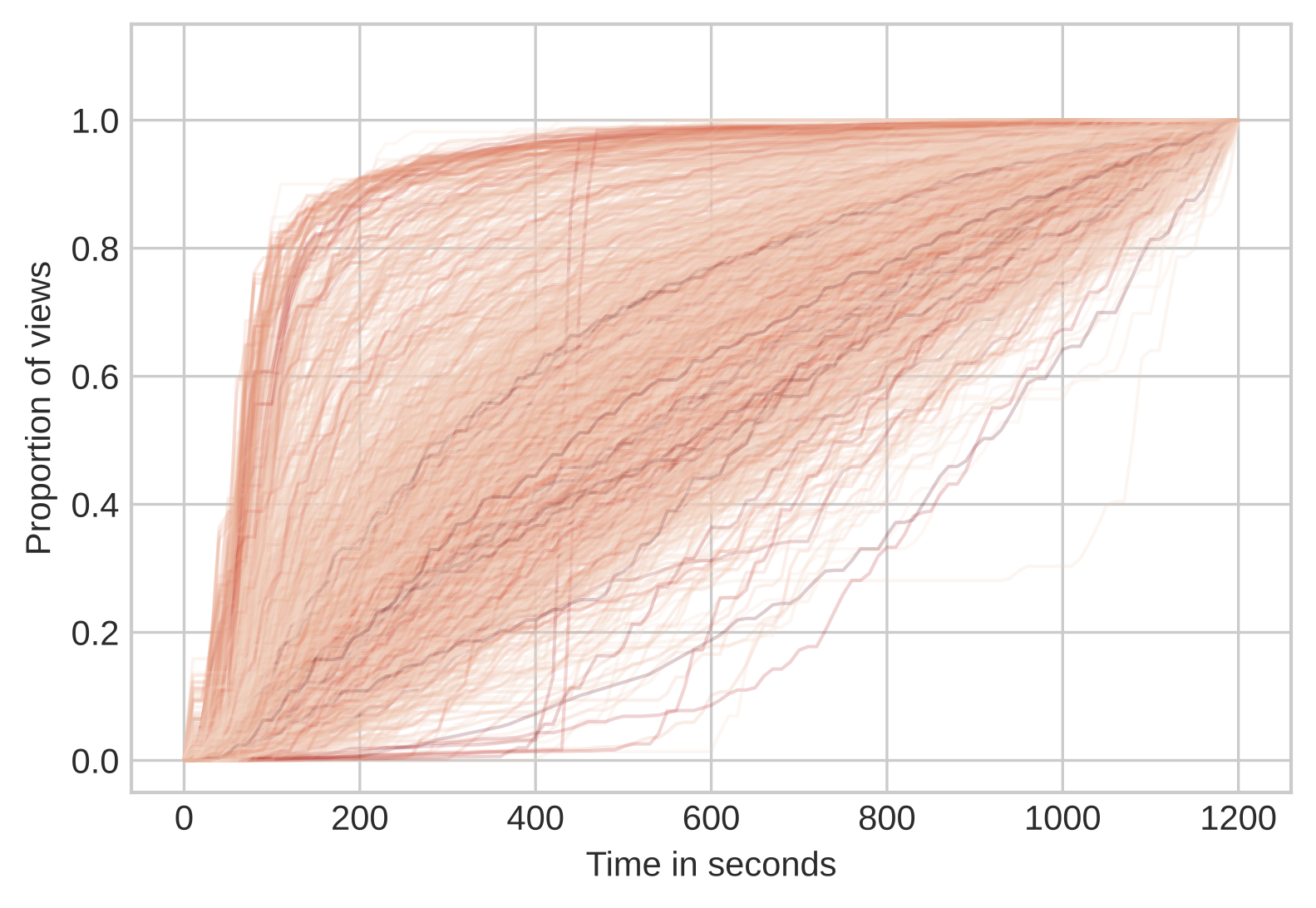}
  \captionof{figure}{Proportion of views over time.}
  \label{fig:curves}
\end{minipage}%
\begin{minipage}{.5\textwidth}
  \centering
  \includegraphics[width=0.9\linewidth]{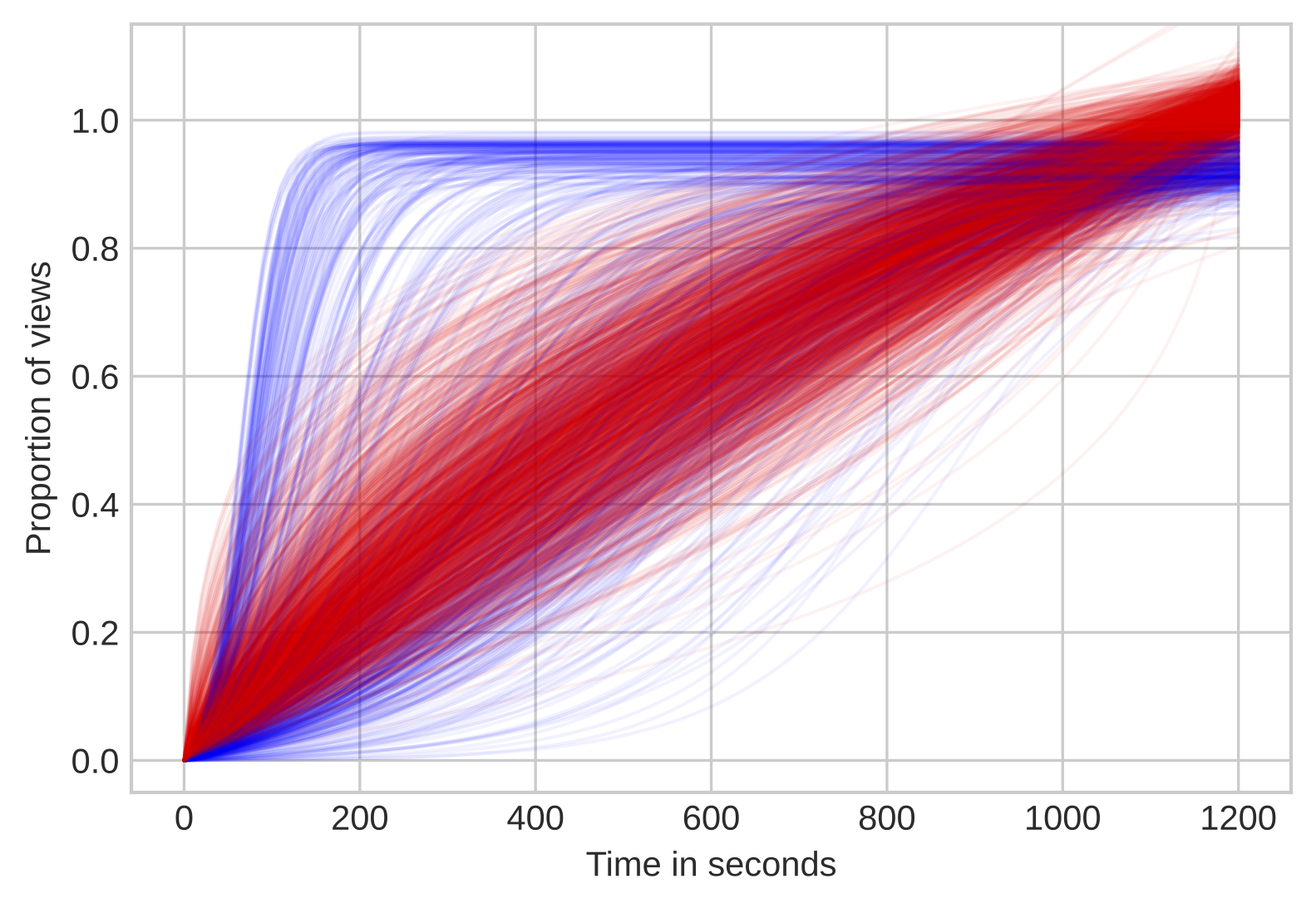}
  \captionof{figure}{Best log (red) and sigmoid (blue) fit over time.}
  \label{fig:curvesfitted}
\end{minipage}
\end{figure*}

\paragraph{Number of Views.}
The time series of the Tweet views were collected on average over 1,893 seconds (${\sim}31.5$ minutes) after the Tweets were sent. During this time, the Tweets accumulated, on average, 46.2 views (range 0--43,870). 15.6\% of Tweets had zero views. Fig. \ref{fig:views} plots the $log_{10}$--distribution of the number of views. Due to the nature of the long-tailed distribution with a small number of Tweets with a very large number of views (about 0.5\% having more than 1,000 views), the median of 7 will better represent the view distribution. 

\paragraph{Diffusion patterns.}
In Fig. \ref{fig:curves}, we have plotted the views-over-time curves for all 2,723 Tweets of our sample that received more than 50 impressions by the end of the data collection. Looking more closely at our data, it becomes obvious that we can observe two different diffusion dynamics.

The \emph{sigmoid} type (eq. 1) represents Tweets that reach their maximum potential for impressions very fast and quickly saturate. Without further analysis, we can assume that these Tweets remain within their local areas of the network and they receive few or no retweets. 

\begin{equation}
T_{a, b}^{sigmoid}(t) = \frac{1}{1 + (b * exp(-a * x))} - \frac{1}{1 + b}    
\end{equation}

\begin{figure*}[t]
\centering
\begin{minipage}{.475\textwidth}
    \centering
    \includegraphics[width=.82\linewidth]{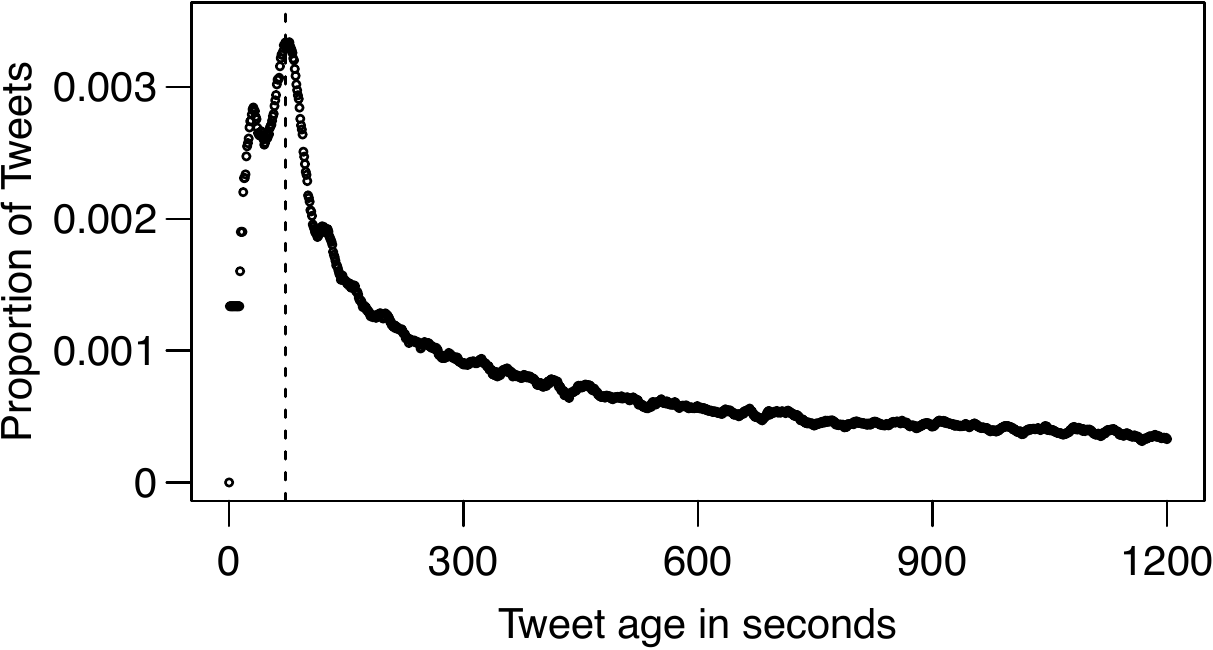}
    \captionof{figure}{Average views per second within the first ${\sim}20$ min.}
    \label{fig:seconds}
\end{minipage}
\hfill
\begin{minipage}{.475\textwidth}
    \centering
    \includegraphics[width=.87\linewidth]{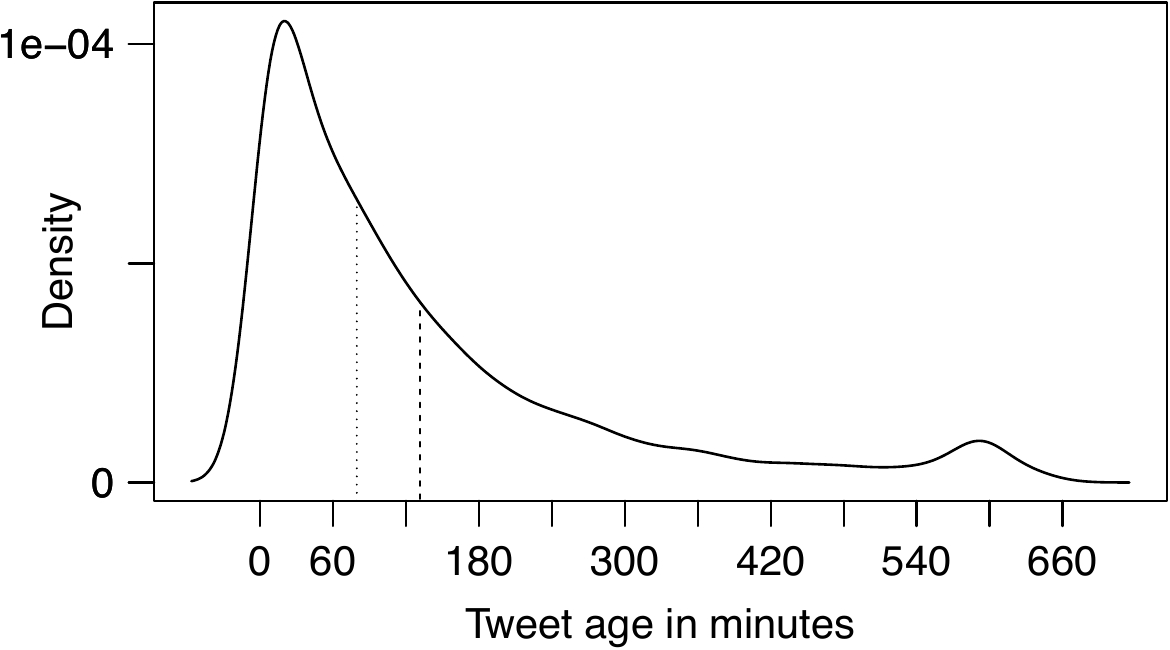}
    \captionof{figure}{Distribution of half-life values with mean = 131.6 min. (dashed line) and median = 79.5 min. (dotted line).}
    \label{fig:halflife}
\end{minipage}
\end{figure*}

Fig. \ref{fig:seconds} (which we will discuss later), appears to imply that new views are distributed according to ${\sim}t^{-1}$, implying the cumulative view count to follow a \emph{log}-curve, which can be described as:

\begin{equation}
T_{a, b}^{log}(t) = b * \frac{log(a * x + 1)}{log(a + 1)}
\end{equation}

Fig. \ref{fig:curvesfitted} is based on the same data as Fig. \ref{fig:curves}. Here, every time series was fit with both model types and drawn with the better fitting (as measured by the MSE every ten seconds) function. The curves are then colored red when the \emph{sigmoid}-model (eq. 1) was used and blue for the \emph{log}-model (eq. 2). The \emph{sigmoid}-model performs better if we allow estimation over 1.0, which makes sense when imagining the future development of the curves.

Identifying the diffusion type of a Tweet at an early stage can be helpful in predicting its future view count development. For our data, categorizing Tweets as \emph{log}- or \emph{sigmoid}-types improves the prediction of how many views they received after 24 hours significantly. \emph{Log}-tweets receive, on average, 29\% more views after 24 hours, with a significance of $\alpha < 1\%$.

\paragraph{15 Minutes of fame?}
Another question for understanding the diffusion processes of Tweets is to estimate the relative temporal peak, i.e., when will most users see a Tweet? To answer this question, we took the approximately 18,000 Tweets of our collected data that were available in all 99 data collections and had at least 1 view overall. We then extracted for every time series the differences in views among all pairs of consecutive collection time stamps and divided these numbers by the number of views for each Tweet after all collection rounds. Every proportion of increase per time interval was then distributed evenly to all seconds of the respective time interval. In other words, we have summed up the distribution of the proportions of the view increases every second for every Tweet. It is important to note that with this approach, we level out two groups of outliers: a) Tweets with a very high view number because they contribute the same 1.0 as an unimportant tweet, and b) tweets with potentially diverging diffusion curves because their diverging contribution will be negligible for the overall curve.

The result of this process can be seen in Fig. \ref{fig:seconds}. The fact that the peak of average views per second is reached quickly, followed by a steep decay, resonates with our previous analytical steps. Here, we can see that the peak at which most impressions per second are created is, on average, 72 seconds after a Tweet is created.

\paragraph{24 hours later.}
To better understand the diffusion dynamics of Tweets beyond the first minutes, we have collected the above-described Tweets again after 1/2/3 days. Let us first compare the view numbers of the Tweets at the age of 1 and 3 days. These results are very clear and are shown in Tab \ref{tab:days}. In a nutshell, almost 1/3 of all Tweets that have gotten views within the first 24 hours do not receive any more views within the subsequent 48 hours, and only about 1 in 20 Tweets can increase the views by more than 50\% during this time span.\footnote{And just so that it is double-checked as well: No single Tweet had fewer views after three days than it had after one day.}

\begin{table}[h]
\centering
\begin{tabular}{lr}
\hline
\textbf{Increase Views Day 1 $\rightarrow$ 3} & \textbf{Percentage} \\
\hline
+ 0\%                                    & 29.6\%  \\
\textless{}=10\%      & 36.3\%  \\
\textgreater{}10\% … \textless{}=50\%      & 28.6\%  \\
\textgreater 50\%                         & 5.5\%   \\         
\hline
\end{tabular}
\caption{Proportion of Tweets that can/cannot increase view counts from day 1 to day 3 }
\label{tab:days}
\end{table}

The picture does look different when comparing the number of views after the initial 20 minutes of observation with the numbers after 24 hours. Here, the median increase in view count is a factor of 3.75.

 \paragraph{Half-life of Tweets.}
Since the view counts from 20 minutes to 24 hours changed by a factor of 3.75, we cannot observe 50\% of views within this dataset, and this data is not sufficient for empirically measuring the half-life of a Tweet. Consequently, we turned to the second dataset, which includes 1,000 data collections of about 5,000 Tweets over the course of 11 hours. Consistently with the previous data collection, 8.5\% of Tweets had zero views after 24 hours. For the remaining Tweets, we have evaluated how long it took for every Tweet to reach 50\% of the 24h view numbers. In less than 4\% of Tweets, this was not possible, i.e., the Tweets reached the 50\% level after the first eleven hours, confirming our previous observation that view counts reduce quickly over time for the vast majority of Tweets. 

Fig. \ref{fig:halflife} illustrates the distribution of half-lives in our second dataset. The right-skewed distribution has an arithmetic mean of $131.6$ minutes (dashed line) and a median of 79.5 minutes (dotted line) with the following quantiles:

\begin{table}[h]
\begin{tabular}{lrrrrr}
Quantile  & 10\% & 25\% & 50\% & 75\%  & 90\%  \\
Half-Life & 7.2  & 26.3 & 79.5 & 175.5 & 342.1
\end{tabular}
\end{table}

\section{Outlook}
\paragraph{Future research questions}
The most obvious future research questions are related to identifying the factors that drive view counts and half-life. At every data collection, we also get the number of re-tweets and likes at the moment of data collection. Mathematically modeling and studying the temporal interplay of these times series with the number of views is a topic for a separate paper. Other features that are available via the Twitter API are the number of followers of the tweet senders, the tweet content, and possibly connected images and websites, to name just a few.

We are aware that there are Tweets that go viral days or even months after they were sent. We did not account for these dynamics. However, long-term phenomena could be studied with our approach of repeatedly collecting information about the same set of Tweets (e.g., once a day).

Finally, studying human behavior with social media data always comes with challenges related to biases and data quality \cite{Ruths2014}. The addition of the impression count to the list of variables, which researchers can get from API calls, will open up great new research opportunities to study popular users and content as well as more nuanced diffusion processes. At the same time, research also has to focus on revealing technical details and possible artifacts of view counts and, more broadly, Twitter metrics.

\paragraph{Secrets.}
One surprising observation of this study was that a significant proportion of Tweets do not get any views. Are these Tweets getting banned, but not deleted? This and many other questions are related to the fact that social media platforms, including Twitter, are secretive about their algorithms and data handling. Besides investigating platform dynamics to improve research quality, we also need to hold the platforms accountable whenever possible to increase transparency about data handling and algorithmic content filtering.

\section{Research Ethics and Reproducibility}
In this study, we used only publicly available data from Twitter and only utilized Twitter's own APIs to collect data. We did not send any Tweets and did not interact with other Twitter accounts. Our only variables extracted from the Twitter data were Tweet IDs, timestamps of when the Tweets were created, and the impression count, which is part of the public metric variable. No Tweet texts, account profile information, or other information that could identify individuals or groups (PII) were analyzed.

\paragraph{Reproducibility.} All data from the analyses of this article are available online (\emph{www.pfeffer.at/data/halflife}). The data includes all Tweet IDs, Tweet creation time, and for each collection iteration for every Tweet, its collection time, and the number of views. Since the views are a function of when the Tweets are collected, we have expanded the JSON response data from the Twitter API that is stored in files with the exact time of every API query.

\section{Acknowledgments}
J.P. wants to thank Dr. Kathleen M. Carley and Dr. Larry Richard Carley for discussing historical literature related to the topic as well as the potential mathematical operationalization of information half-life. 

\bibliography{aaai23}

\end{document}